\documentclass[twocolumn,fleqn,usenatbib]{aastex631}

 \UseRawInputEncoding
\usepackage{enumitem}

\usepackage{amsfonts}
\usepackage{amsmath}
\usepackage{amssymb}
\usepackage{multirow}   
\usepackage{bbm}
\usepackage{cleveref}
\usepackage{multirow}
\usepackage{pgfplots}
\usepackage{placeins}
\usepackage{tikz}
\usepackage{upgreek}
\usepackage{booktabs}
\usepackage[normalem]{ulem}

\usepackage{ulem}
\usepackage{cancel}
\usepackage{graphicx}
\usepackage{color}

\usepackage[T1]{fontenc}

\DeclareRobustCommand{\VAN}[3]{#2}
\let\VANthebibliography\thebibliography
\def\thebibliography{\DeclareRobustCommand{\VAN}[3]{##3}\VANthebibliography}

\newcommand{\be}{\begin{equation}}
\newcommand{\ee}{\end{equation}}
\def\lsim{\mathrel{\rlap{\lower4pt\hbox{\hskip0.5pt$\sim$}}
    \raise1pt\hbox{$<$}}}         
\def\gsim{\mathrel{\rlap{\lower4pt\hbox{\hskip0.5pt$\sim$}}
    \raise1pt\hbox{$>$}}}         

\def\la{\langle}
\def\ra{\rangle}
\def\dd{{\rm d}}
\def\ln{{\rm ln}}

\def\lsim{~\rlap{$<$}{\lower 1.0ex\hbox{$\sim$}}}
\def\bsim{~\rlap{$>$}{\lower 1.0ex\hbox{$\sim$}}}
\def\kms{\ {\rm km\,s^{-1}}}

\def\kpc{\ {\rm Kpc}}
\def\mpc{\ {\rm Mpc}}
\def\pc{\ {\rm pc}}

\def\vx{\mathbf{x}}

\def\apjs{{Astrophys.~ J.~ Suppl.~}}
\def\apjl{{Astrophys.~ J.~ Lett.~}}

\def\Sg{\Sigma_{g}}

\def\Sast{\Sigma_{\ast}}

\def\Sacr{\Sigma_\textrm{acc}}
\def\Sacr{\Sigma_\textrm{acc}}
\def\Macr{M_\textrm{acc}}
\def\dSacr{{\dot \Sigma}_\textrm{acc}}
\def\fsf{f_{SF}}

\def\mpc{\, \rm Mpc}
\def\msol{\, {\rm M}_\odot}
\def\myr{\, {\rm Myr}}
\def\gyr{\, {\rm Gyr}}

\def\kms{\, {\rm km }\, {\rm s}^{-1}}
\def\kpc{\, {\rm kpc }}

\def\lcdm{$\Lambda$CDM}
\def\mhone{$\mh=10^{11}\msol$}
\def\mhtwo{$\mh=10^{12}\msol$}
\def\mh{M_\textrm{h}}
\def\dmh{{\dot M}_\textrm{h}}
\def\dmst{{\dot M}_*}
\def\fsf{f_\textrm{SF}}
\def\fb{f_\textrm{b}}
\def\mv{M_\textrm{h}}

\def\rv{r_\textrm{h}}
\def\rd{R_\textrm{d}}
\def\fb{f_\textrm{b}}
\def\vc{V_\textrm{c}}
\def\lB{\lambda_\textrm{B}}
\def\rpt{\textsc{VELOCIraptor}}

\def\muv{M_\textrm{UV}}
\definecolor{RedWine}{rgb}{0.743,0,0}
\definecolor{RoyalBlue}{rgb}{0.25,.41,.88}
\definecolor{ForestGreen}{rgb}{.13,.54,.13}
\definecolor{DeepPurple}{rgb}{.72,.18,1}
\pgfplotsset{compat=1.18} 
\begin{document}

\title{High-redshift halo-galaxy connection  via constrained simulations}

\author[0000-0002-8272-4779]{Adi Nusser}   
\email{adi@physics.technion.ac.il}
\affiliation{ Department of Physics and the Asher Space Research Institute\\ Israel Institute of Technology Technion, Haifa 32000, Israel}



\begin{abstract}

The evolution of halos with masses around $M_\textrm{h} \approx 10^{11}\;  \textrm{M}_\odot$ and $M_\textrm{h} \approx 10^{12}\;  \textrm{M}_\odot$ at redshifts $z>9$ is examined using constrained N-body simulations. {The average  specific mass accretion rates, $\dot{M}_\textrm{h} / M_\textrm{h}$, exhibit minimal mass dependence and generally agree with existing literature.  Individual halo accretion histories, however, vary substantially. } About one-third of simulations reveal an increase in $\dot{M}_\textrm{h}$ around $z\approx 13$. 
Comparing simulated halos with observed galaxies having spectroscopic redshifts, we find that for galaxies at $z\gtrsim9$, the ratio between observed star formation rate (SFR) and $\dot{M}_\textrm{h}$ is approximately $2\%$. This ratio remains consistent for the stellar-to-halo mass ratio (SHMR) but only for $z\gtrsim 10$. At $z\simeq 9$, the SHMR is notably lower by a factor of a few.
At $z\gtrsim10$, there is an  agreement between  specific star formation rates (sSFRs) and $\dot{M}_\textrm{h} / M_\textrm{h}$. However, at $z\simeq 9$, observed sSFRs exceed simulated values by a factor of two.
  {It is argued that the mildly elevated SHMR }  in high-$z$ halos with $M_\textrm{h} \approx 10^{11} M_{\odot}$, can be achieved by  assuming the applicability of the local Kennicutt-Schmidt law
 and  a reduced effectiveness of stellar feedback due to deeper gravitational potential of high-$z$  halos of a fixed mass.
\end{abstract}


\section{Introduction}

The standard \lcdm\ cosmological model, incorporating a cosmological constant, $\Lambda$, and cold dark matter (DM), has been remarkably successful in interpreting and predicting fundamental properties of the large-scale structure of the Universe. Despite potential tensions \citep[e.g.][]{kids21,riessJWST2023}, this success extends to temperature anisotropies of  the cosmic microwave background (CMB),  clustering of  the distribution of galaxies, and  deviations of  galaxy motions from a purely Hubble flow
\citep[e.g.][]{ColeBAO2005,Eisenstein2007,Davis2011,carrick15,PlanckCollaboration2018, LilowNusser2021}. 
On galactic scales, predicting the properties of the galaxy population and its evolution with redshift has been less straightforward. This complexity arises from the intricate nature of baryonic physics involved in star formation processes, including gas dynamics, heating and cooling mechanisms, and notably, the energetic feedback from supernovae (SN) and active galactic nuclei (AGNs)  \citep[e.g.][]{Larson74,DS86,White1991,Silk1998,okalidis21,Krumholz2018a,NS22}.  

Prior to the era of the James Webb Space Telescope (JWST) \citep{jwst}, significant efforts have been invested in {developing} models of galaxy formation to adequately describe observations at low and moderately high redshifts ($z\lsim 10$) \citep[e.g.][]{finkelstein_hubble_2022}.

Observations obtained with the JWST have significantly deepened our view of the universe, revealing galaxies as far back as a couple of hundred million years near the Big Bang. However, the JWST has also detected an unexpected excess of luminous galaxies at higher redshifts. While the initial findings from the JWST {appeared to pose}  serious  challenges for the standard \lcdm\  model of structure formation, the severity of these discrepancies were {significantly}  alleviated with more precise calibration and the availability of spectroscopic redshifts \citep[cf.][for an overview]{yung24}.

It should be emphasized that the star formation rates (SFRs) in high-redshift JWST galaxies are not particularly unusual in themselves  \citep[e.g.][]{robertson23,harikane24}. These galaxies exhibit SFRs that can be adequately sustained by cosmological gas accretion onto halos \citep{Mason2023}. Matching the abundance of halos  to  the observed distribution of UV magnitudes (used as proxies for the SFRs) of galaxies at $z\gsim 10$ implies that these galaxies should be hosted in halos of mass $\mh \approx 5\times 10^{10}-10^{11}\msol$ \citep{kolchin23,Mason2023,ChenMo23}. For such halos, the star formation efficiency $\fsf$ (i.e. the fraction of accreting gas turning into stars) needed to account for the SFRs, is $\gtrsim 0.13$ (see   \ref{s:hah} below). 
At low redshifts ($z\lesssim 4$),  the stellar-to-halo mass ratio (SHMR) 
inferred from abundance matching is in the range $0.001-0.01$ for $\mh \approx 10^{11}\msol $ halos \citep[e.g.][]{Moster2013,Rodriguez-Puebla2017,girelli20,Fu2022}.   
Assuming a global gas fraction $\fb=0.157$ in galaxies \citep{PlanckCollaboration2018}, this implies 
an average  star formation efficiency, $\fsf\approx 0.06 -0.006$, which is at least a factor of two lower than
the inferred value at $z\gtrsim 10$.


 {An important aspect of star formation inside DM halos is their mass accretion history \citep[e.g.][]{White1991}. Halo accretion is directly linked to the availability of gas for star formation. Newly accreted gas replenishes the reservoir, which is subsequently converted into stars and may escape the galaxy through processes like SN and AGN feedback. In this paper, we assume that $z\gtrsim 10$ galaxies indeed inhabit massive halos and aim to numerically investigate the assembly history of these halos. 
}
 Numerical studies of  individual  objects typically rely on the methodology of zoom-in simulations \citep[e.g.][]{zoomin, auriga, sun2023a,pallottini23}. In this type of simulations  high resolution is employed only  in a small region allowing a detailed study of its  small scale dynamics while  simultaneously 
capturing the interaction within the  larger cosmic environment.
In this paper we invoke an alternative approach  of constrained simulations \citep[e.g.][]{Romano-Diaz2005} to model the high redshift $z>9$ 
accretion history of halos above $\mh \approx 10^{11} \msol$. 
We utilize the \citet{hr91} method 
of constrained random realizations to generate initial conditions that are guaranteed to contain a halo in specified mass range, when evolved forward to a specified redshift, $z$.

The technique of constrained simulations is very useful in large scale structure studies, especially for assessing uncertainties in  parameter estimations realistically and in mitigating  cosmic variance \citep{Hellwing2017}. This approach  has been used to derive simulation  initial conditions from the observed peculiar velocities \citep{hoffman15} and from the 2MRS redshift survey \citep{LilowNusser2021}.
Initial conditions based on the galaxy distribution in the Sloan Digital Sky Survey (SDSS)  survey have also been
generated to run a constrained simulation to study the local Universe \citep{WangMo16}. The same simulation has also been utilised to study the 
$z\simeq 0$ descendants of galaxies at $z\approx 8-9 $ galaxies \citep{ChenMo23}.

The structure  of the paper is outlined as follows. In \S\ref{s:hmf}, we assess the abundance of halos and compare is with 
the luminosity function of galaxies at high redshifts. We underscore the advantage of  constrained simulations based on the expected abundance of the host halos. 
 The simulations are detailed  in \S\ref{s:sim}, which includes the  method for generating suitable initial conditions. Additionally, this section contrasts the accretion history of halos in the simulations with observational data. In \S\ref{sec:model}, a straightforward  recipe for star formation is introduced. The recipes produces a  higher ratio of stellar to halo mass at high redshifts compared to low redshift. A  summary and discussion are provided  in \S\ref{sec:disc}.

We adopt the standard flat \lcdm\ cosmological model \citep{PlanckCollaboration2018} with a total mass density parameter $\Omega_m=0.311$, baryonic density $\Omega_b=0.049$, Hubble constant $H_0=67.7\kms\, \mpc^{-1}$ and normalization $\sigma_8=0.81$.

\section{Halo abundance at high redshift}

\label{s:hmf}

 We employ a halo definition in terms of a spherical overdensity, where the halo virial radius $\rv(t)$, at any given time $t$ is determined such that  the mean density within this radius equals $200$ times the critical density of the universe, $\rho_c(t)= 3H(t)^2/8\pi G$. Therefore,
\begin{align}
\label{eq:vir}
\rv(t) & =  0.1 H(t)^{-1}V_c \; , \nonumber \\
\mv(t) & =  0.1 G^{-1} H(t)^{-1} V_c^3 \; , 
\end{align}
where the
 circular velocity $\vc=\sqrt{G \mv/\rv}$.
 At high redshifts where $H\sim 1/t$, these relations yield 
 \begin{align}
\label{eq:virN}
\rv(t) & =  13.1\kpc \left(\frac{11}{1+z}\right)^{3/2}\frac{V_c}{181\kms} \; , \nonumber \\
\mv(t) & =   10^{11}\msol \left(\frac{11}{1+z}\right)^{3/2}\left(\frac{V_c}{181\kms}\right)^3 \; , 
\end{align}

\begin{figure}
\hskip -0.4cm
\includegraphics[width=.5\textwidth]{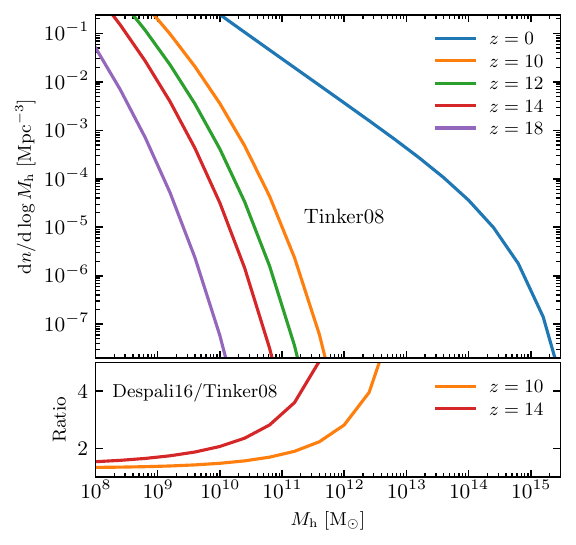} 
\vskip 0.1cm
\caption{\textit{Top:} Halo abundance versus mass at different redshifts as denoted  in the figure. The solid lines are derived from the {Tinker08} expression for the mass function in the Planck \lcdm\ cosmology (see text).
At the high mass end, the curves are $\sim \mh^{-4}$ and thus the cumulative number density $n(>\mh) \approx (1/4) \dd n/\dd \log \mh$. \textit{Bottom:} Ratio of {Despali16}  to {Tinker08} HMFs.}
 \label{fig:nm_colussus}
\end{figure}

\Cref{fig:nm_colussus} displays the abundance of DM halos per logarithmic {mass}  bin per 
$\mpc^3$ at different redshifts. 
The plots are generated using  the widely used halo mass function (HMF)  outlined by \citet{Tinker08} (hereafter Tinker08) for the $\Lambda$CDM cosmological model, as incorporated within the COLOSSUS cosmology Python package \citep{colossus}. The curves corresponding to halos within the mass range $\mh \gsim 10^{11} \msol$ at redshifts $z \gsim 9$ exhibit comparable abundance to large groups and clusters at $z=0$.
 At the upper end of the mass range, the dependence on mass steepens significantly  at such high redshifts. For instance,  at $z=9$, the abundance of halos with $\mh =10^{12} \msol  $  is four orders of magnitudes 
lower that than halos with $\mh=10^{11} \msol $.
A reasonable approximation to the HMF is given by 
\begin{equation}
\label{eq:tinker_app}
\frac{\dd n(z,\mh)}{\dd \log \mh} = \frac{10^{-5}{\mpc}^{-3}} {\left[\left(\frac{\mh}{M_{-5}(z)}\right)^2+0.25 \left(\frac{\mh}{M_{{-5}}(z)}\right)^{0.8}(z)\right]^2}\; .  
\end{equation}
The approximation is valid in the redshift range  $ 10<z<20 $ with $M_{-5}(z)$ is defined as,
\begin{equation}
    \log \left(M_{-5}/10^{11}\msol\right)=  0.185(z-10)^{1.03} \; , 
\end{equation}
and is equal to the halo mass where $\dd n/\dd \log M =10^{-5}\mpc^{-3}$.
For $M\gsim M_{-5}(z)$, we have the steep dependence  $\dd n/\dd \log M\sim M^{-4}$.

Various fitting formulae for  the mass function are available in the literature \citep[e.g.][]{press74,sheth99,Jenkins2001,angulo12,seppi21}. 
Therefore, it is important  to investigate whether discrepancies among these mass functions might be particularly notable  when applied to high redshifts.
The bottom panel of \Cref{fig:nm_colussus} compares the Tinker08 with {another} widely used HMF given in 
 \citet{despali16} (hereafter Despali16).
 The ratio between these HMFs increases with mass and redshift but remains within a factor of a few.
Further, due to the steep dependence of the number density on halo mass, {  we shall see in \cref{fig:Mh_MUV} that differences in the halo mass function lead to minor changes in the estimation of halo masses by matching UV LF.}

\subsection{Abundance matching}
\citet{harikane24} (hereafter H24) constrain the UV LF of high-$z$ galaxies using 25 galaxies with spectroscopic redshifts spanning $z\approx 8.61-13.20$. Their constraints align with various luminosity distributions derived from photometric redshifts \citep[e.g.,][]{perez23, donnan23, harikane23, bouwens23}. H24 show that the observed UV LF can be effectively modeled by a double power-law function, denoted as $\Phi_\textrm{UV}$. 

\begin{figure}
\hskip -0.4cm
\includegraphics[width=.5\textwidth]{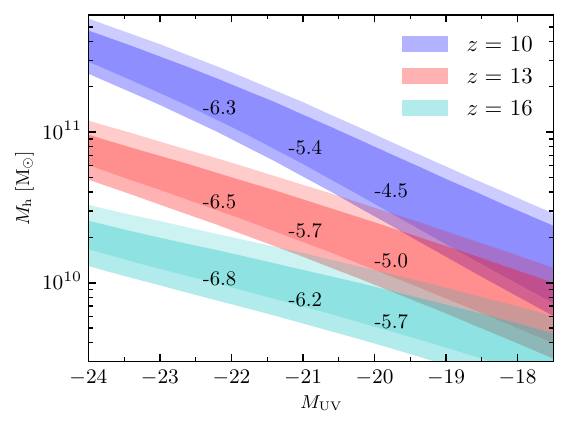} 
\vskip 0.1cm
\caption{Halo mass versus UV magnitude obtained by abundance matching of 
the {Tinker08} (dark shaded areas) and {Despali16}  (light shades) halo mass function to  the double power law fit of H24 to  UV LF, $\Phi_\textrm{UV}$. Each shaded area corresponds to  a single redshift and is bounded by $\mv$ curves obtained using $5\Phi_\textrm{UV}$ (yielding the lower boundary of the shaded area) and by $0.2\Phi_\textrm{UV}$ (leading to the upper boundary). The numbers in each shaded area represent  the log of the cumulative halo abundance (in $\textrm{Mpc}^{-3}$)
expected at the corresponding $\muv$.
}
 \label{fig:Mh_MUV}
\end{figure}

 Using this double power-law fit, we conduct a straightforward abundance matching to associate galaxies with a UV magnitude, $M_\textrm{UV}$, to halos of mass $\mv$
The results of are summarized in \cref{fig:Mh_MUV} 
for three redshift values and for the HMFs  of {Tinker08} (dark shaded area) and {Despali16} (light shaded). The width of each shaded area corresponds to  variation in the normalization  of the double power law fit, spanning  a factor in between $0.2 $ and $5$. While this should provisionally reflect  the uncertainty in the measured UV LF of H24, it is important to note that the uncertainty in the  observed  luminosity  in a single bin of one UV magnitude  width could be be as large as two orders of magnitude.
{The results for the Tinker08 and {Despali16} HMFs  are remarkably consistent. This consistency is due to the steep dependence of the mass function on the mass of rare halos, approximately $\sim \mh^{3.5-4}$, which implies weak sensitivity of $\mh$ to the observed number of galaxies in a given UV bin.}

At $M_\textrm{UV} =-21 $, the halo mass is in the range 
$\mh=6\times 10^{10} - 1.5\times 10^{11}\msol$ at $ z=10$, with a 
 cumulative halo  abundance (indicated by log the number density in the shaded areas) similar to massive groups and clusters  at $z=0$.  Although galaxies with  a fixed $\muv$ correspond to lower $\mh$ as we move from low to high redshifts, the decrease in $\mh$ is insufficient  to maintain the same abundance. In fact, halos corresponding to a  fixed $\muv$ become rarer.
 
At  $z\approx 10$, on average  a single halo with  $\mh =2\times 10^{11}\msol $ is
expected in a box a $100\mpc$. While it is  possible to employ zoom-in techniques for simulating massive halos at $z\gsim 10$, this necessitates significant computational resources. Instead, we adopt a computationally friendlier approach, utilizing constrained random realizations to generate initial conditions that are guaranteed to contain a massive halo when evolved forward to a specified redshift, $z$.

\section{Simulations}

\label{s:sim}

In \ref{s:constr}, we outline the method for generating constrained initial conditions. Using these initial conditions, simulations of DM particles were conducted in periodic boxes using the SWIFT cosmological code \citep{swift23}. All simulations started at redshift $z_{i}=80$ and concluded at $z=9$. As we shall see below, the linear density contrast corresponding to the constraint is $\delta_c=1.68$ at $z=9$, and hence the corresponding density at the initial redshift $z_i$ is $0.02$, well within the linear regime. Furthermore, to capture any mild deviations from linear evolution at $z_i$, the initial conditions are generated using the Zel'dovich approximation, rather than linear theory. 

We performed  nine simulations constrained to have halo masses of approximately $\mh \approx 10^{11}\msol$ in boxes of $L=17.1 \mpc$, along with one unconstrained simulation in a box of the same size. Additionally, two simulations were conducted with initial conditions constrained to include  a halo mass of $\mh \approx 10^{12}\msol $ at $z=9$, positioned at the center of cubic boxes of $L=36.9 \mpc$.

Each simulation consisted of $512^3$ equal-mass particles, resulting in particle masses of $10^7\msol$ and $10^6 \msol$ in the large and small boxes, respectively. The maximum physical softening used in the simulations was $100\pc$. Throughout each simulation run, the output of positions and velocities of all particles was retained at 18 different redshifts spanning from $z=20$ to $z=9$. Halos were identified from the simulation outputs utilizing the \rpt\ halo finder \citep{elahi19}. This halo finder provides  halo masses according to several definitions. Here we use \rpt\ masses  that match the definition in  \cref{eq:vir}.

\subsection{Constrained Initial Conditions}
 \label{s:constr}

We formulate the condition  (constraint) for the presence  of a halo of a given mass as follows. 
Let $\delta(\vx,z)$ be the linearly evolved density at redshift $z$ and 
$\delta_R(\vx,z)$ be its convolution with a top-hat window of  comoving radius $R$. We associate a halo of mass $\mh$ with a  comoving Lagrangian  radius $R_L=(3\mh/4\pi\rho_\textrm{m})^{1/3}$, where $\rho_\textrm{m}$ is the background density in comoving coordinates. 

The formation of a halo of 
 mass $\mh$ at redshift $z$, located at position   $\vx_0 $  is determined by the condition $\delta_{R=R_L}(\vx_0,z)=\delta_c\simeq 1.68$ \citep[e.g.][]{PS,Peebles1980TheUniverse}.  Here, $\delta_c$ is the critical threshold indicative of the  virialization of DM halos. 
 At $z=10$, halos with masses \mhone\ and \mhtwo\ correspond to Lagrangian comoving radii of $R_L=0.84 \, \textrm{Mpc}$ and $1.82 \, \textrm{Mpc}$, respectively. In terms of the ratio $\delta_c/\sigma_{R_L}$, these are equal to  $4.8$ and $6.5$, providing a measure of   halo formation  likelihood  under the specified conditions.
 This formulation  is only approximate as 
the superposition  of generic fluctuations on all scales and non-linear evolution will  lead to deviations from the desired halo mass and position.
Nonetheless,  the prescription is reasonable for massive (rare) halos \citep{robertson09,ludlow14}, as is the case in the current study.

We adopt the methodology of  \citet{hr91} to generate gaussian random fields that satisfy  the aforementioned condition. This  method 
expresses the constrained random field, denoted by $\delta_1$,  in terms of an unconstrained random gaussian field, $\delta^\textrm{unc}$, as follows 
\begin{equation}
\label{eq:hr1}
\delta_1(\vx,z) = \delta^\textrm{unc}(\vx,z)-\langle \delta(\vx,z)|{\delta_{R_L}^\textrm{unc}}\rangle+\la  \delta(\vx,z)|{\delta_c}\ra\; 
\end{equation}
 where  ${\delta_{R_L}^\textrm{unc}} $ is the value of the filtered unconstrained field at $\vx_0$.
The ensemble average of all $\delta$ fields satisfying the constraint $\delta_{R_{L}}(\mathbf{x}_0,z)=C$ is given by
\begin{equation}
\label{eq:HR}
\langle \delta(\mathbf{x})|C \rangle = \zeta(|\mathbf{x} - \mathbf{x}_0|) \frac{C}{\sigma_{R_L}^2}\; ,
\end{equation}
where $\sigma^2_{R_L}$ is the variance of the smoothed density, $\delta_{R_L}$.

\begin{figure}
\includegraphics[width=.5\textwidth]{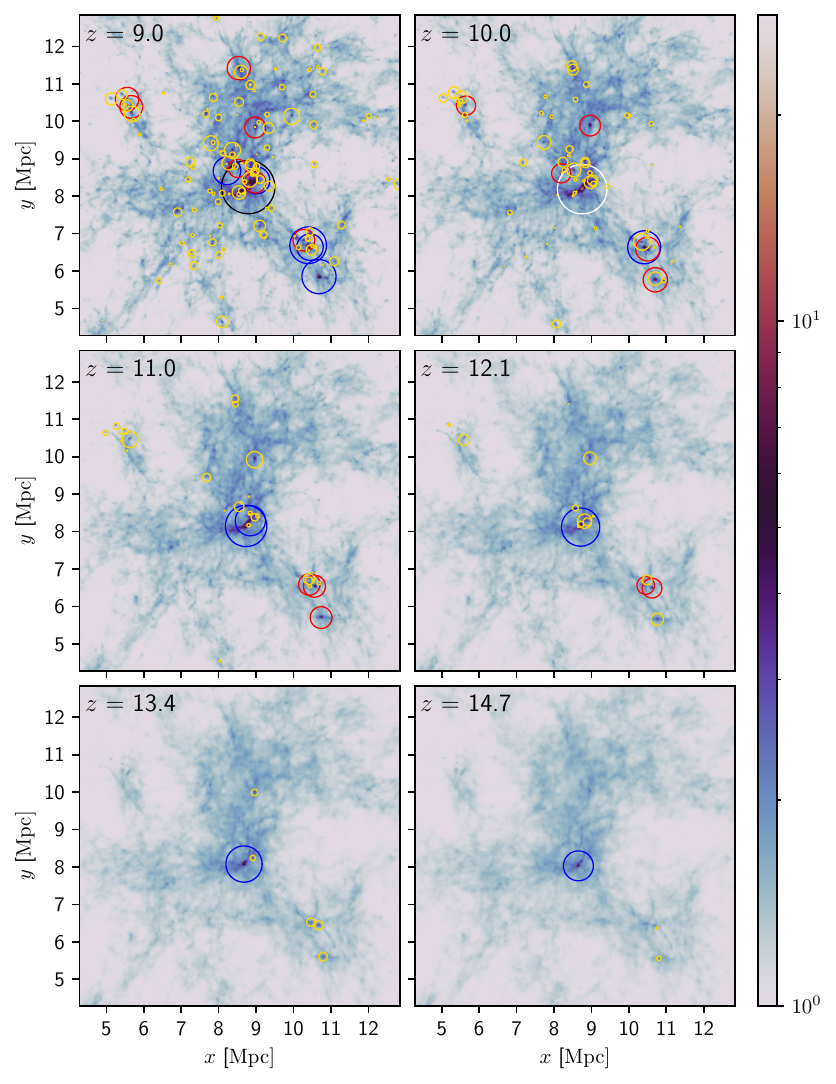} 
\vskip 0.5cm
\caption{Projected 2D density field  from  one of the constrained  simulations 
with  $L=17.1 \mpc$ at six distinct redshifts.
Circles denote halos classified by mass ranges: black ($\mh/\msol >10^{11}$), white ($5\times 10^{10} - 10^{11}$), blue ($10^{10} - 5\times 10^{10}$), red ($5\times 10^{9} - 10^{10}$), and yellow ($10^{9} - 5\times 10^{9}$), with size logarithmically scaled by mass. The color bar indicates density  relative to the mean density within the simulation box.}
 \label{fig:snap}
\end{figure}

\begin{figure}
\includegraphics[width=.49\textwidth]{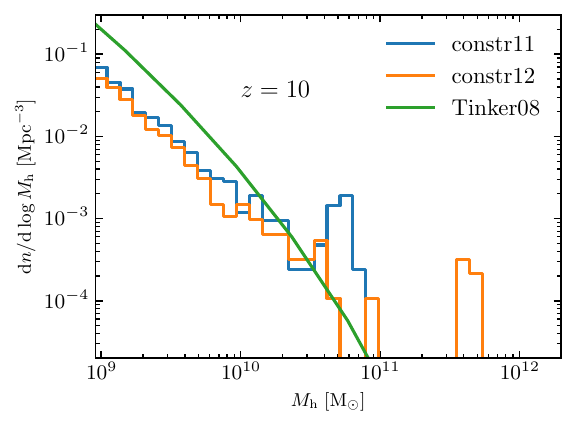} 
\vskip 0.cm
\caption{Halo abundance computed from the simulations output at $z=9$. The blue and orange histograms represent the constrained simulations with halo masses of $10^{11}, M_{\odot}$ and $10^{12}, M_{\odot}$, respectively. The green curve depicts the {Tinker08} mass function.}
 \label{fig:nm_simu}
\end{figure}

\subsection{Results}
\label{s:results}

\Cref{fig:snap} illustrates the projected (2D) density   (in units of the mean 2D density value) for one of the \mhone\ constrained  simulations. The overlaid circles represent identified halos in different mass ranges, as described in the figure caption. Only halos above $10^9\msol$ are marked. 

The displayed region of the box is focused around the center.  As expected, the most massive halo (MMH) forms
near the center. Additional massive halos are associated with the growth of the MMH, however, their masses are significantly lower than the  MMH. Halos in the mass range $5\times 10^9<\mh /\msol <10^{10} $ (red circles) are present at $z\simeq 12$. These halos are not considered rare, as their expected abundance is 
$10^{-2}-10^{-3} \textrm{dex}^{-1} \mpc^{-3} $ (see \cref{fig:nm_colussus}). 
Therefore, we expect to identify a few 
such halos even in our simulation box of $L=17.1 \mpc$.

Inspection of the panels at $z=11$ and $z=10$ reveals a major merger event with the two halos marked by two blue circles ($10^{10}<\mh/\msol<5\times 10^{10}$) at $z=11$ merging to form the larger halo indicated by the white circle ($5\times 10^{10}<\mh /\msol <10^{11}$) at $z=10$. There is a clear tendency of increasing halo mass as we move nearer to the MMH, particularly  at the lowest redshifts.

 \Cref{fig:nm_simu} shows the abundance of halos from the simulations as a function of $\mh$. At the high mass end, the simulated abundance significantly exceeds the predictions from the {Tinker08} fitting formula (green curve). In both \mhone\ and \mhtwo\ constrained simulations, the MMH is accompanied by other relatively rare massive halos with abundance well above the green curve. {The initial condition constraint ensures a massive halo in a small region, making the simulation box atypical for its volume. Consequently, the number density of massive halos in the simulation significantly exceeds the expected mean, explaining the enhanced abundance of high $\mh$ halos compared to {Tinker08}. One pathway for the formation of these massive halos is through the accretion of smaller halos, leading to a modest depletion in the low $\mh$ range. This explains why the simulated halo abundance at low $\mh$ falls slightly below the green curve, typically by a factor of $3-4$.}

\subsubsection{Halo Accretion History}
\label{s:hah}

\begin{figure}
\centering
  \includegraphics[width=0.95\linewidth]{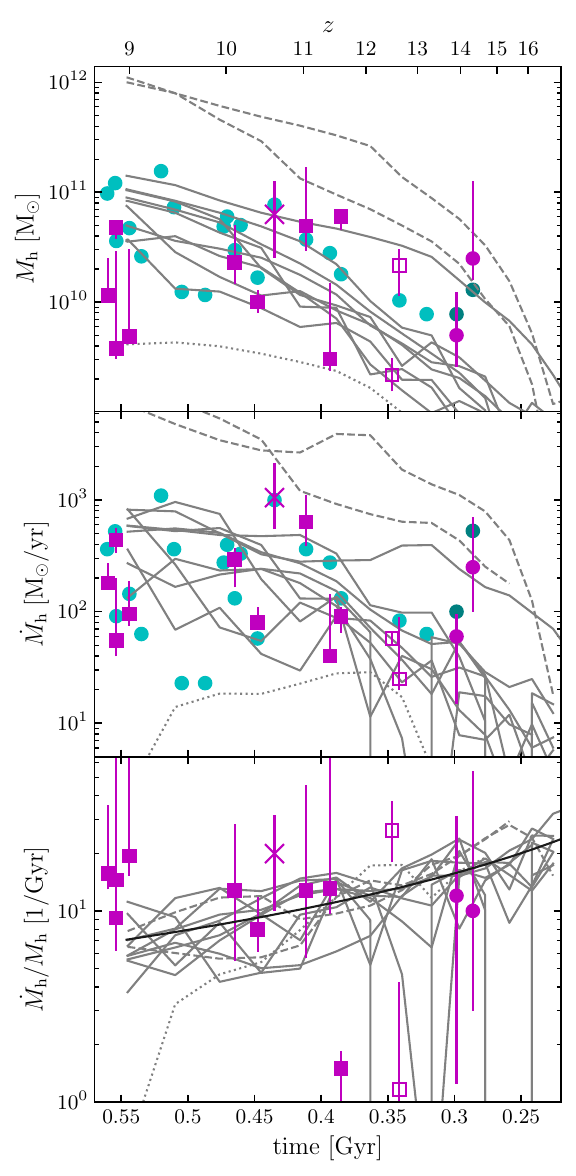}
\caption{Accretion histories of main halos in unconstrained (dotted) and constrained simulations with $M_h=10^{11}\msol$ (solid) and $10^{12}\msol$ (dashed). Observational data shown as magenta and cyan symbols. \textit{Top:} Halo mass vs. time. Magenta: derived from SED-fitted $M_\ast$ in the literature, assuming $\mh=50M_*$  (see  \cref{eq:mst_to_mh}). Cyan: from abundance matching of observed $M_\textrm{UV}$. \textit{Middle:} Accretion rate $\dot{M}_h$. Magenta: $\dmh=50\dmst$ from observed SED-fitted $\dmst$. Cyan: from UV-estimated $\dot{M}_\ast$. \textit{Bottom:} Curves of specific accretion rate $\dmh/\mh$, with the fit  $\sim t^{-4/3}$  plotted in black, and observed sSFR $\dmst/M_\ast$ shown as magenta points. See text for details.}
\label{fig:mzt}
\end{figure}

\Cref{fig:mzt} displays the evolution of MMH   properties in simulations (continuous curves) and compares them with observational data (individual symbols). The figure is divided into three panels, each focusing on a different aspect of halo growth.
The data symbols are as follows.
\setlength{\leftmargini}{0.5cm} 
\begin{itemize}

\item \textit{Magenta symbols  with error bars} are based on stellar masses and SFRs inferred via SED fitting.
   The \textit{filled squares} are taken from table 3 in H24.
The \textit{open squares} refer to the galaxy GS-z12 for which  different values  of redshift,   SFR and $M_\ast$ are reported in \citet{eugenio23} and H24. Both sets of data are shown, with \citet{eugenio23} being  the point with the lower redshift ($z=12.43$).
 {The \textit{cross} is the galaxy GN-z11 \citep{GNZ11}. The two highest redshift points represented by \textit{filled circles} correspond to the two galaxies reported in \citet{Carniani2024}. 
}

The observed stellar mass, $M_*$, and SFRs, $\dmst$, are used to estimate halo masses and mass accretion rates assuming 
  assuming
\begin{equation}
\label{eq:mst_to_mh}
\mh=\fb^{-1} \fsf^{-1} M_\ast= 50\left(\frac{0.13}{\fsf} \right) M_*\; ,
\end{equation}
and similarly for the relation between $\dmh$ and $\dmst$. As before, $f_b$ is the global barynonic mass fraction, and  the star formation efficient $\fsf$ is a constant  assigned a default value $\fsf =0.13$, implying $\mh=50 M_*$
\item \textit{Cyan circles } are based solely on the observed  $\muv$ provided in   table 1 of H24. The \textit{dark cyan circles} correspond to the two highest redshift galaxies, with $\muv$ taken from  \citet{Carniani2024}.
For these points, halo masses are inferred from $\muv$ via abundance matching from the observed UV magnitudes. From \cref{fig:Mh_MUV}, the relative uncertainty in these points is a factor of $\approx 2-3$, but we do not attach the corresponding error bars for the sake of clarity.  
The SFRs are deduced directly using  ${\dot M}_\ast(\textrm{M}_\odot \, \textrm{yr}^{-1})= 1.15 \times 10^{-28} L_{UV}(\mathrm{erg\, s^{-1} \,Hz^{-1}}$ assuming a Salpeter IMF. The halo 
accretion rate is them estimated from \cref{eq:mst_to_mh}.

\end{itemize}

 \noindent \textbf{Top panel: halo mass vs. time.}
{The grey curves correspond to  $\mh(t)$ of the simulated MMHs. The curves  } reveal that half of the simulated MMHs, including the unconstrained halo (dotted), have acquired 80\% of their final masses in the last $150 \myr$. Only the dashed curves, corresponding to constrained simulations with $\mh \approx 10^{12}\msol$ and one of the nine solid curves, have acquired a mass $\gtrsim 10^{10} \msol $ by $z=15$.

The \textit{Cyan circles } fall within the range of solid curves corresponding to simulations constrained to contain a $\mh\approx 10^{11}$ halo. 
{The result is not entirely trivial since $\mh$ of the MMHs  been tuned to match the abundance at redshifts   $z\approx 9$ and not at higher redshifts. Indeed, at higher redshifts,  the spread in halo masses between different simulations is becomes large, ranging from $\mh \approx 10^{10}\msol$ to $10^{11}\msol$ even at at $z=10$, } as indicated by the solid curves. This mass range is associated with more than a two-order-of-magnitude difference in the abundance of halos, as shown in  \cref{fig:nm_colussus}.

The highest redshift data point represented by a  \textit{magenta circle} is well  above all  solid curves except  one (yellow curve). Nonetheless, since we have only nine curves corresponding to the \mhone\ simulations, we conclude that this data point is consistent with simulated accretion history.

 At $z\gsim 10$, the  estimates from \cref{eq:mst_to_mh} (magenta)  agree with both the \mhone\ simulations and abundance matching results (cyan circles). However, at $z\simeq 9$, these estimates fall below both simulations and abundance matching.   This is in agreement   with various models in the literature  \citep[e.g.][]{behroozi15,Mauerhofer23,harikane24,yung24} predicting an LF  consistent with  the observations  at $z\lsim 9$, but underestimating the observed abundances at $z\gsim  11$.

\bigskip 

\noindent \textbf{Middle panel: mass accretion rate.} As in the previous panel, the curves correspond to the simulations. The $\dmh$ curves reveal significant variations between individual halos. Some halos exhibit highly fluctuating $\mh$, while others (e.g., those represented by dashed and a few solid curves) show smoother evolution. However, even these smoother curves display fluctuations on timescales $\lsim 100\myr$.

The magenta and cyan points agree with each other, but are not identical, as the SED fitting involves more detailed SFR modeling than UV magnitudes alone. Accretion rate curves from the \mhone\ simulations are consistent with observations via $M_\textrm{UV}$ (cyan) and SED-SFRs (magenta).

 We emphasize that while both cyan and magenta points in top and middle panels rely on $\muv$, their methodologies differ: abundance matching for the top panel versus an empirical SFR-$\muv$ relationship for the middle panel.

\bigskip

\noindent \textbf{Bottom panel: specific accretion rate.}
The specific halo accretion rates, $\dmh/\mh$, from simulations cluster  around a simple fit denoted by a black line, represented by the equation ${\dd \ln \mh}/{\dd t} = 3.15 t_{\textrm{Gyr}}^{-4/3} \gyr^{-1}$. This fit approximates the mean accretion rate for halos of mass $\mh=10^{11}\msol$ as proposed by \citet{fakhouri10}.

The magenta points represent the sSFR, $\dmst/M_*$, from observations. Cyan circles are absent from this panel since only $\dmst$ can be directly derived from $\muv$.

 At $z>10$, there is a reasonable agreement  between the observed  sSFRs   and   the  halo  accretion rates from the simulations. 
 This consistency   corroborates the findings of \citet{bouwens23}, who noted   that the sSFRs tend to follow  the scaling $(1+z)^{2.5}$ proposed  by \citet{fakhouri10} for the specific halo accretion rate  at high redshift. However,  the normalization of the specific halo  accretion rate in  \citet{bouwens23} is higher by a factor of a few compared to  the fit by \citet{fakhouri10}.

At $z<9$, the discrepancy between data and observations seen on the top panel is also evident  here. 
While reducing $\fsf$ by a factor of $\approx 5-10$ (from $\fsf=0.13$ to $0.01-0.025$) would reconcile the $z\simeq 9$ with the $\mh$ results in the top panel, it would not effect  the halo specific accretion rate
(assuming a constant $\fsf$).  The is because, according to  \cref{eq:mst_to_mh}, the specific accretion rate is equal to the sSFR, i.e. $\dmh/\mh=\dot M_\ast/M_\ast$, independently of $\fsf$. 

Note that $M_\ast\propto \mh^\alpha$ ($\alpha=const$), then $ \dmh/\mh=\alpha^{-1} \dot M_\ast/M_\ast$ \citep{behroozi15}.
 Therefore, also for $\alpha\ne 1$  curves of the sSFR and  the specific halo accretion rate should trace each other, with a constant ratio between them. Thus, a mass-dependent $\fsf$ would not resolve the discrepancy.


\section{Boosting the star formation efficiency at high-$z$ }
\label{sec:model}

A boost in $\fsf$ to $\approx 0.13$ at high-$z$ yields reasonable agreement between simulated halos and observations, representing a mild increase relative to low-$z$. We present 
a simple recipe explaining this enhancement, suggesting that star formation processes may not significantly differ across redshifts. We present a model explaining this enhancement, arguing that star formation processes may not significantly differ across redshifts.
For a halo of mass $\mh(t_z)$ at redshift $z$, we model $M_\ast$ and $\dot M_\ast$ assuming star formation occurs in a rotationally supported disk governed by the Schmidt-Kennicutt (SK) law \citep{Schmidt1959,Kennicutt1998}:
\begin{equation}
\label{eq:SK}
{\dot \Sigma}_\ast  =A \Sg^n ; ,
\end{equation}
where ${\dot \Sigma}_\ast$ is the SFR per unit disk area, $n=1.54$, and $A=10^{-3.95}$ \citep{Kennicutt2021}. $\Sigma_\ast$ and $\Sg$ are in $\textrm{M}_\odot ,\textrm{pc}^{-2}$.
Disk gas partially converts to stars following the SK law, with stellar feedback expelling a fraction. The disk gas reservoir is simultaneously replenished and expanded through halo accretion.

The gas surface density evolution is described by,
\begin{equation}
\label{eq:mass_balance}
    {\dot \Sigma}_g = \dSacr- {\dot \Sigma}_\ast -{\dot \Sigma}_\textrm{ej} \; ,
\end{equation}
where $\dSacr$ is accretion, ${\dot \Sigma}_\ast$ is star formation, and ${\dot \Sigma}_\textrm{ej}$ is feedback-driven ejection. For halos with $\mh \lsim 10^{12}M_\odot$, AGN feedback is subdominant \citep{Croton2006,Bower2006,puchwein13}.

Gas ejection is modeled via \citep{White1991,Kauffmann1993,Mitchel2018,yung2019},
\begin{equation}
\label{eq:eject}
   {\dot \Sigma}_\textrm{ej}=  \left(\frac{V_c}{V_{SN}}\right)^{-\gamma}{\dot \Sigma}_\ast
\; ,
\end{equation}
with $V_{SN}=240\kms$ and $\gamma = 2.8$ \citep{yung2019}. This implies more effective stellar feedback in halos with shallow gravitational potential \citep{Larson74,DS86}.

The accreted gas mass in time $\delta t$ is,
\begin{equation}
\label{eq:dmacr}
    \delta \Macr=\fb {\dot M}_\textrm{h}\delta t\; . 
\end{equation}

Due to short crossing times and efficient cooling \citep{dekel23,yung24}, we assume rapid settling into an exponential disk:
\begin{equation}
\label{eq:dsacr}
\delta \Sacr(t,R) = \delta \Sigma_0 e^{-R/R_d(t)}\; ,
\end{equation}
where $\rd = 0.7 \lB \rv$ \citep{mmw,yang23}, and $\lB$ is the halo spin parameter \citep{bullock01}.
Therefore,
\begin{equation} 
\label{eq:dSacr}
\dSacr(t,R) = \frac{\fb \dot \mv }{2\pi \rd^2(t)} e^{-R/\rd(t)}\; .
\end{equation}

Motivated by our simulations, we assume $\dmh/\mh \propto t^{-4/3}$, yielding:
\begin{equation}
\label{eq:mh_fit}
\mh(t)=\mh(t_z) \mathrm{e}^{A(t_z^{-\beta}-t^{-\beta})}; ,
\end{equation}
where $A=-8.1$, $\beta= 1/3$, and $t$ is in Gyr.

We numerically integrate \cref{eq:SK,eq:dSacr} from $t=t_i\ll t_z$ to $t=t_z$. Initial conditions are set as $\Sast(t_i,R)=10^{-3}\Sg(t_i,R)$, with $\Sg(t_i,R)$ following an exponential profile. $\rd(t_i)$ is determined by $\rd = 0.7 \lB \rv$ with $\rv=\rv(t_i)$. The initial disk gas mass equals $\fb \mh(t_i)$ minus the initial stellar mass. We use $\lB=0.035$ in all calculations.

\begin{figure}
\includegraphics[width=.45\textwidth]{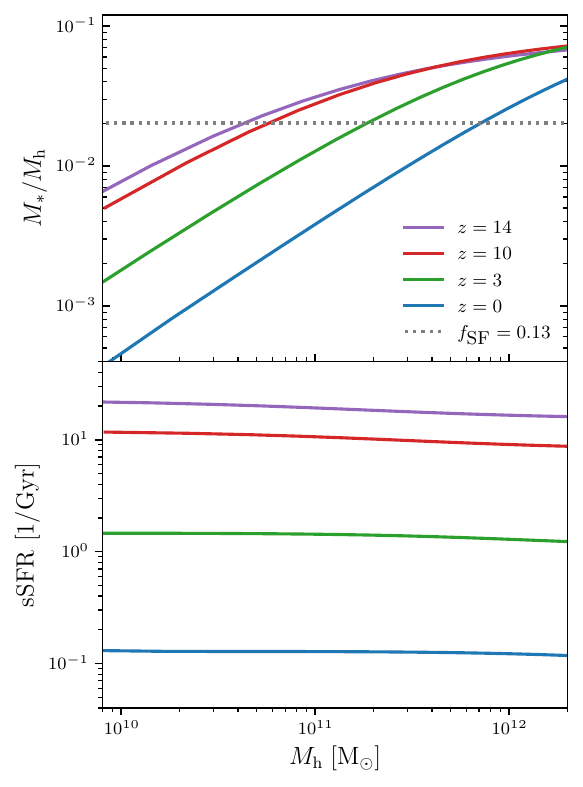} %
\vskip 0.1cm
\caption{The SHMR (top panel) and the sSFR (bottom panel) from  
the simplified recipe in \S\ref{sec:model} at four  different redshifts, as indicated in the figure. The dotted grey horizontal line indicates the value corresponding to $M_*/M_\mathrm{h}=\fsf\fb $ obtained with  $\fsf=0.13$.}
 \label{fig:shmr}
\end{figure}

In  the top panel of \cref{fig:shmr}, the SHMR is plotted against halo mass for four  redshift values, $z$. The model agrees reasonably well with the observed local SHMR as estimated through abundance matching techniques \citep[e.g.][]{Moster2013,Rodriguez-Puebla2017,moster20,shuntov22,girelli20}.

The model SHMR acquires larger values at higher redshifts for a given mass. This is due to the relation $V_c \sim \mv/t$, indicating that a higher circular velocity $V_c$ occurs at earlier times for a fixed $\mv$, leading to less efficient SN feedback and, consequently, a larger gas reservoir for star formation.  The SHMR for $\mh \approx 10^{11} \msol$ increases by approximately a factor of five  at high-$z$ compared to $z=0$, while for $10^{10} \msol$, the increase is about a factor of 25. In contrast, predictions from the \textsc{UniverseMachine} \citep{behroozi20} suggest that the SHMR increases by about a factor of 10 from $z=0$ to $z=12$ for halos with $10^{10} \msol$ (their figure 12).

The  SHMRs  change very little between $z=10$ and $14$ with a weak dependence on $\mh$. For the relevant mass range $\mh \approx 5 \times 10^{10}- 10^{11}\msol $ the SHMR is 
$\approx 2\%$ corresponding to $\fsf\approx 0.13$, the value used in figure \cref{fig:mzt}. 

For $\mh \approx 10^{11}\msol$, at $z=3$,  the SHMR changes by a factor of $\approx 4-5$ compared to $z=0$. This may seem at odds  with observational analyses in the literature, which generally suggest a constant SHMR over this redshift range.  We defer a detailed discussion of this issue  to \S\ref{sec:disc}.

The sSFRs plotted in the bottom panel are close to the observed values at the corresponding  redshifts \citep{bouwens23} and depends weakly on $\mh$.

These results suggest that the enhanced star formation efficiency in high-$z$ galaxies can be explained by the fundamental physics of structure formation and feedback processes, without invoking drastically different star formation mechanisms compared to the local universe. 

\section{summary and discussion}
\label{sec:disc}

We have presented a study of the accretion history of massive halos at redshifts $z\gtrsim$, relevant to luminous galaxies observed at such high redshifts. 
Our approach is based on constrained simulations, which   is highly beneficial  for studying rare cosmological structures. 
Here we have only conducted simulations with constant resolution across the entire simulation box. However, a combination of zoom-in techniques and constrained initial conditions is most appropriate for 
resolving rare structures as well as capturing the gravitational influence of the large scale environment.

Growing evidence suggests highly variable star formation history at high redshifts \citep{Cole2023CEERS:Bursts,dressler24}, potentially due to mergers, interactions, and environmental conditions. This variability could bias inferred UV luminosity distributions  \citep{Ren2019,Mason2023,sun2023,sun2023a,Shen_Vogl2024}, as galaxies in low-mass halos may be preferentially detected during increased star formation phases. {In the simulations, individual halo accretion curves exhibit both long-term fluctuations ($\gsim 100 \myr$) and short-term variations. Examining \cref{fig:mzt} (middle panel) reveals a tendency for greater variability in lower mass halos at $z=9$ compared to more massive ones, potentially leading to enhanced stochasticity in associated star formation rates. However, our output times do not capture variability at $\lsim 10 \myr$ scales. }

{ High-resolution simulations yield mixed results: SERRA simulations do not produce sufficient star formation rate variability \citep{pallottini23}, while FIRE-2 simulations show bursty star formation that explains the observed UV luminosity function (LF) \citep{sun2023a}. However, stellar masses at $z\gsim 10$ in these simulations \citep{MaHopkins2018} are lower than observed estimates from spectroscopically confirmed galaxies \citep{harikane24}.}
 Nonetheless, stochasticity is clearly an important effect that should be considered.

In \cref{fig:mzt} we have seen that  dividing the observed SFRs by a factor $f_b \fsf =2\% $ (i.e. $\fsf=0.13$), leads to $\dmh $ that are consistent with the simulations constrained to include  a halo of mass  $ \mh \approx 10^{11} \msol$. The agreement spans  the entire  considered redshift range, $z>9$.
Interestingly, dividing the  observed stellar masses by the same factor  yields a good match with the halo masses in the simulations, but this is only true for  $z\gtrsim 10$. For galaxies at $z\simeq 9$, the factor required is smaller by a factor of $\approx 5-10$  ($\fsf\approx 0.025-0.01$), closer to what is seen in low redshift galaxies. 

This peculiar behaviour of the inferred $\mh$ between $z\approx 9$ and $z\approx 10$ may  stem from challenges  in accurately estimating the stellar masses. 
Indeed,  SFRs estimated from UV magnitudes are generally more reliable  than stellar mass estimates, which require assumptions about the entire star formation history \citep[e.g.][]{whitler23} SED fitting \citep{Dayal2023,pallottini23,whitler23,pacifici23,wangB23}. A striking example is the galaxy GS-z12 ($z=12.48$). Its estimated mass varies by an order of magnitude depending on the method used:  $M_\ast = 4.3^{+1.8}_{-2} \times 10^8 \msol$ using PROSPECTOR \citep{PROSPECTOR,harikane24} and $M\ast = 4.36^{+1.8}_{-1.27}\times 10^7 \msol$ using BEAGLE \citep{BEAGLE,eugenio23}.
Nonetheless, in  the estimating the SFRs, the impact of potential uncertainties due to  potential  dust attenuation needs to be assessed 
 \citep[e.g.][]{Kennicutt2012}. However, dust attenuation is expected to be small in these high redshift galaxies \citep[e.g.][]{bouwens23}, suggesting that the observed discrepancies could primarily be due to the complexities of stellar mass estimation.

Another possibility for this behaviour  is an abrupt 
change in the conditions for star formation at $z\approx 9-10$, similar to the  suggestion of \citet{sbnnw23} although their model refers to transition at $z\approx 6 $.

Numerical simulations are computationally intensive for tracing the accretion history of a large ensemble of halos. Semi-analytic methods for generating constrained merger histories \citep{Nadler2023} offer a CPU-efficient alternative. These could be valuable for exploring variations in past accretion rates of rare halos at high redshifts. Currently, these methods have been applied to trace the growth of rare halos from $z\approx 12$ to $z=0$, rather than tracing rare halos at $z\approx 10$ backward in time.

Numerous models aim to understand the formation of luminous high-$z$ galaxies. \citet{harikane23} suggested that UV-inferred SFRs might be overestimated due to a top-heavy IMF at high redshift, which could arise naturally in low-metallicity environments. \citet{yung24} noted this could account for a factor of 4 boost in UV luminosities, aligning their semi-analytic models with observations.
\citet{dekel23} propose conditions for feedback-free star formation in $\approx 10^7\msol$ gas clouds at high redshifts, satisfied in $10^{11}\msol$ halos at $z\approx 10$. 
Conversely, \citet{sbnnw23} invoke AGN positive feedback, suggesting short-lived AGN activity triggers vigorous star formation via momentum-conserving outflows. They predict a transition to energy-conserving flows at $z\approx 6$, leading to gas depletion and quenched star formation at lower redshifts.
\citet{ferrara23} propose that decreased dust attenuation at high redshifts could explain the abundance of $z\gtrsim 10$ galaxies, compensating for reduced host halo abundance.
{Modifications to the primordial mass power spectrum have also been explored \citep{Padman2024,Laha2023,hirano2024,munoz2024}.}

In the approximate star formation recipe outlined in \S\ref{sec:model}, we used the halo's circular velocity, $V_c$, as the parameter governing SN feedback. 
Combined with the local KS law, the recipe aims to demonstrate that  negative feedback at high-$z$ is naturally expected to be less efficient than at lower $z$.

According \cref{fig:shmr} the recipe  implies an evolving SHMR at moderate redshifts and a non-evolving one at $z\gtrsim 10$. 
Observationally, within the uncertainties, a non-evolving SHMR is generally consistent with galaxy luminosity distributions up to $z \lsim 5$. However, for the redshift range $z \approx 0 - 10$, studies in the literature show divergent results. Some find weak to moderate redshift dependence \citep{Mason2015,Rodriguez-Puebla2017,Moster2013,moster20,stefanon21}, while others report significant evolution \citep{Behroozi2013,Sun_Fur2016, Behroozi2019}. Notably, substantial differences exist between various SHMR estimates at similar redshifts and halo masses.

The evolution of the SHMR, inferred through abundance matching techniques, is sensitive to various factors, including the shape of the galaxy stellar mass function \citep{yang2012}. As highlighted by \citet{Fu2022}, uncertainties in observed stellar mass functions at redshifts $z\lesssim 4$ can lead to differing interpretations regarding the evolution of the SHMR from $z=0$ to $z=4$. Depending on specific assumptions about the stellar mass function, the SHMR could exhibit either a decrease or an increase over this redshift range. Depending on the assumptions, the SHMR for a halo with $\mh \approx 10^{11} \msol$ could vary by two orders of magnitude due to different assumptions. 

The recipe in  \S\ref{sec:model} can be adapted to yield a nearly non-varying SHMR at $z\lesssim 5$ through several modifications.
For example, the adopted expression for $\mh(t)$ is approximate and neglects variations between halos, which can be significant according to \cref{fig:mzt}. The expression  motivated by our high-$z$ simulation results. A recipe with slower accretion at moderate redshifts would yield a less varying SHMR at $z\lesssim 5$. 

Furthermore, following \citep{yung2019}, we have adopted a gas ejection expression determined by $V_c$. However, the maximum circular velocity (the peak of the rotation curve), $V_{\textrm{max}}$, is likely more relevant since it better reflects the depth of the gravitational potential of the halo.

In the regime of stable clustering, $V_{\textrm{max}}$ is expected to remain constant over long cosmic epochs. Thus, employing $V_{\textrm{max}}$ instead of $V_c$ in the model should result in a more constant SHMR over extended periods. We have run the recipe with $V_\textrm{max}$ instead of $V_c$ in \cref{eq:eject}, where the dependence on $\mh$ and $z$ follows the formula obtained by \citet{puebla16} by fitting the median growth of halos in MultiDark N-body simulations. This has yielded closer curves for the SHMR at $z=0$ and $z=3$, while leaving the curves at $z=10$ and $14$ virtually unchanged.

\section{Data availability}
{New numerical simulation data  have been generated and analyzed.}

\section{acknowledgements}
The author has benefited from fruitful conversations with Andrew Benson, Enzo Branchini, Stephane Charlot, Avishai Dekel and Joe Silk. 
This research is supported by a grant from the Israeli Science Foundation and a grant from the Asher Space Research Institute.
The research in this paper made use of the SWIFT open-source simulation code
(http://www.swiftsim.com, \cite{swift18}) version 1.0.0.



\bibliography{references.bib}{}
\bibliographystyle{aasjournal}
\end{document}